# The altmetric performance of publications authored by Brazilian researchers: analysis of CNPq productivity scholarship holders

**Ronaldo Ferreira de Araujo**
0000-0003-0778-9561
Federal University of Alagoas, Brazil
ronaldfa@gmail.com

**Marcelo Alves**
0000-0003-4995-6612
Fluminense Federal University, Brazil
marceloalves.ufsj@hotmail.com

**Abstract:** The present work seeks to analyse the altmetric performance of Brazilian publications authored by researchers who are productivity scholarship holders (PQ) of the National Council of Scientific and Technological Development (CNPq). It was considered, within the scope of this research, the PQs in activity in October, 2017 (n = 14.609). The scientific production registered on Lattes was collected via GetLattesData and filtered by articles from academic journals published between 2016 and October 2017 that hold the Digital Object Identifier (n = 99064). The online attention data are analysed according to their distribution by density and variation; language of the publication and field of knowledge; and by average performance of the type of source that has provided its altmetric values. The density evidences the long tail behavior of the variable, with most part of the articles with altmetrics score = 0, while few articles have a high index. The average of the online attention indicates a better performance of articles written in English and belonging to the Health and Biological Sciences field of knowledge. As for the sources, there was a good performance from Mendeley, followed by Twitter and a low coverage from Facebook.

**1 Introduction**

The attention economy has been considered an adequate approach to the information management which considers the human attention a scarce merchandise, requiring the application of the economic theory to solve the problems of informational management. This perspective places attention as a strategic resource of the human action, and faced with the abundant and immediate growth of the available contents, it becomes a limiting factor in the consuption of information.

The attention is the mental engament focused on a given piece of information. The items come into our conciousness, we attend to a specific item among many of them and then we decide to act (Davenport & Beck, 2001). Thus, attention is used to filter the most relevant information from a large set of information that involves the human being and proliferates more and more in the networks and social media of digital environments.

It is in this context, in the field of scientific communication and the recognition of the need for filters to assist researchers to make a decision about dedicating to one item to the detriment of another, since "nobody can read everything", that Priem et al. (2010), on their manifesto, emphasize the need for more tools and research based on altmetrics.

The altmetrics has been considered a field of study that is dedicated to the use of cybermetric data for scientometric analysis (Gouveia, 2017) and is born with the possibility of rapid measurement of online attention indicators that products of science achieve shortly after their publication. Therefore, the value of its indicator seems to express the online attention through dissemination and other types of interaction that such products generate in the web, reaching not only the interested people inside the scientific community, but also the general public – whose appropriation indicates social impact.

Although social media are announced as major democratizers of scientific knowledge, the level of this democratization is closely related to aspects of the geopolitical economy of each device in the scientific communication, and as it can vary its membership in different countries and users profiles, it can also present indicators of varied attention regarding the areas of knowledge they cover and the particularities they present.

This current work aims to analyze the altmetric performance of publications authored by brazilian researchers that are productivity in research scholarship holders from the National Council for Scientific and Technological Development (CNPq). This kind of scholarship is aimed at researchers who stand out among their peers, valuing their scientific production according to normative criteria (CNPq, 2017), being attributed to researchers from all fields of study, based not only on the quality of the submitted project, but also on the "quality' of the researcher (Wainer & Vieira, 2013).

There are research studies dedicated to the evaluation of scientific production resulted from research funded with productivity in research scholarships, such as Wainer and Vieira (2013). Nevertheless the focus is generally on the analysis of traditional metrics and bibliometric measures. Therefore, it is necesary to focus on studies that look at the same production, but with a view not only on the academic impact they receive, from the citations, but also on the online attention that such research arouse inside and outside the scientific community.

**2 Material and method**

The database of this research was composed by a sample of Brazilian researchers, identified by a web scraping application, with the extraction of information from the CNPq website of all Productivity Scholarship Holders in Brazil in October of 2017, resulting in a total of n = 14.609. Their publication registered in the Lattes Curriculum were then tabulated from the GetLattesData package (Perlin, 2017). The publications were filtered by academic journal articles published between 2016 and October of 2017 that have the Digital Object Identifier (DOI) infomed on Lattes, n = 99064. Lastly, duplicated DOIs were excluded, referring to articles in co-authorship, resulting in a final sample of n = 69419.

The extraction of online attention data in this production was perfomed by the rAltmetric (rOpenSci, 2017) package that consults the public Application Programming Interface (API) of Altmetrics using the DOI of the publications and returns a spreadsheet with the data produced by the platform. After that, the metadata provided by CNPq (membership, level of the Productivity in Research Scholarship, Large Area and Area) and by Lattes (Language) were crossed in order to produce a single spreadsheet with all the information relevant to this work.

It is important to notice that the Attention Score Altmetric is a general measure of the attention that one article or data set has received online and must reflect the visibility (reach), the influence and engagement (quality of the attention, location posted and reputation of who posted) that can vary according to the source or type of media considered (Altmetric, 2017).

The online attention data are analysed according to their distribuition by density and variation, language of the publication, the article field of knowledge, and the average performance according to the type of source that has provided their altmetric values.

# 3 Analysis and discussion of results

The available solutions for monitoring and analysis of online attention and altmetric data of the scientific production require from this production identification standards, such as DOI, that guarantee, among other advantages, its distinction. The first characterization of the publications to be observed is that 73,41% of it has attribution to the identifier. This seems to indicate that the productivity scholarship holders are keen to publish their research results in journals that have this standard.

The distribution of the variable score (online attention) according to its density and variation can be seen in Figure 1. In the first case, it is noticed that the density evidences the long tail behavior of the variable, that is, most part of the articles do not have online attention, which means score altmetrics = 0, while very few articles have a high index.

**Figure 1.** Distribution of online attention by degrees of density and variation

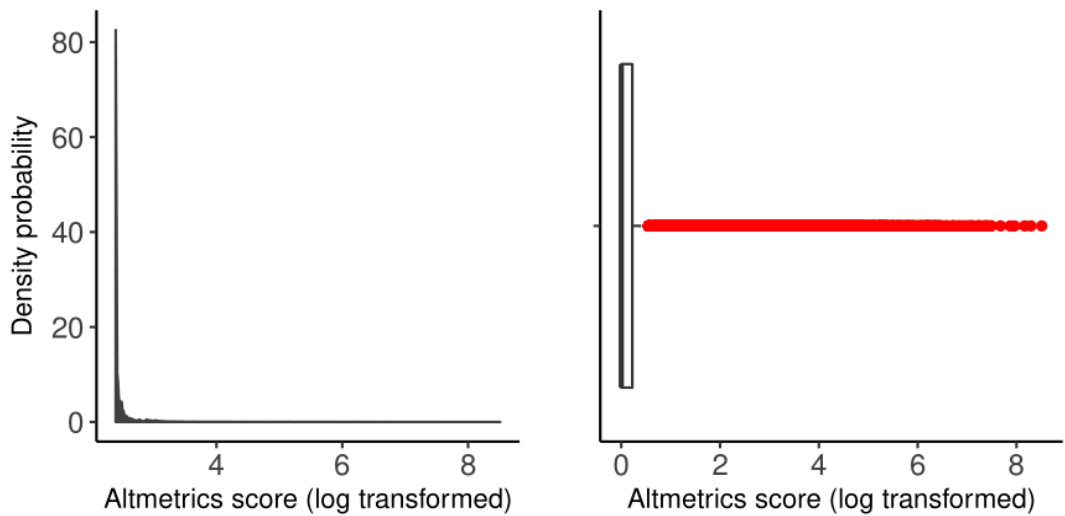

Source: research data (2017)

The second case reinforces the behavior of the former, the data show that all the measures of position are close to 0 and the variable score has a very large number of outliers, ranging from 0 to 4966.648. The average distribution of online attention by language (Graph 1) and area of study (Graph 2) were also observed, considering that they characterize the performance of the analysed production.

**Graph 1.** Average of online attention by language

**Graph 2.** Average of online attention by area of study

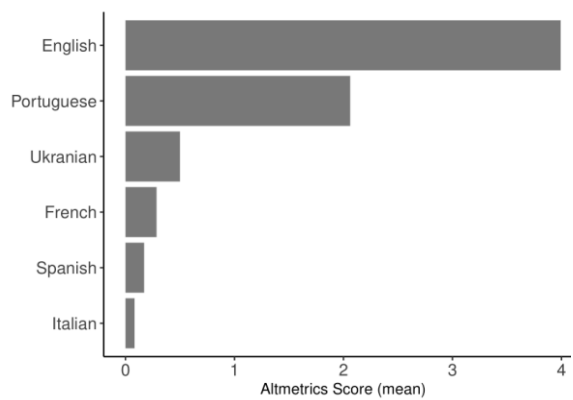
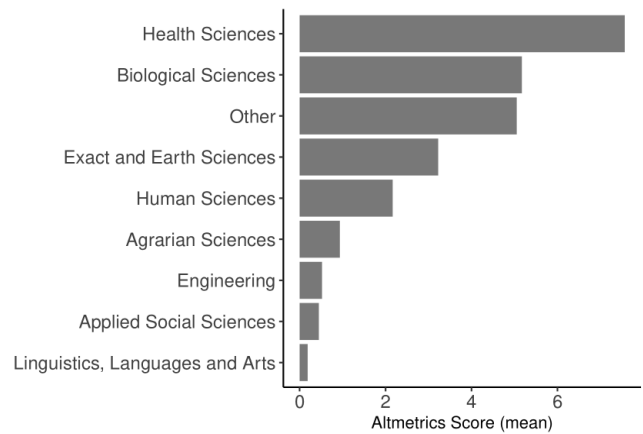

Source: research data (2017)                         Source: research data (2017)

Contrary to the notion of democratization of scientific knowledge attributed to social media, some research show that often the analisys of its use reproduces the same visibility disparities and traditional metrics, such as the fact that publications from emerging countries tend to have less visibility on social media platforms (Alperin, 2015) and that English is the dominant language (Fausto et al., 2012). In the latter case, the result obtained, with its representation of the average of online attention by language, in Graph 1, only confirms such behavior. As for Graph 2, the average of online attention indicates a better performance of Health Siences, with 7.56 and Biological Sciences, with 5.17 points. The result is similar to the studies of Costas, Zahedi and Wouters in which the Health and Biological Sciences are also presented as areas of study that have received greater online attention, with altmetric values higher than the others. But in general, the values obtained by the production of Brazilian researchers are very low.

The *Attention Score Altmetric* is composed by more than 10 distinct sources such as news portals (press), blogs, social media, reference managers, among others. Table 1 presents the average distribution of online attention of the production analysed by the most expressive sources (Mendeley, Twitter, Facebook, Press and Blogs). Sources such as Wikipedia, YouTube, Reddit had a very low value of incidence and were disconsidered.

| Area of study | Twitter | Facebook | Mendeley | Media | Blogs |
|---|---|---|---|---|---|
| Agrarian Sciences | 3.85 | 1.55 | 10.45 | 3.50 | 1.37 |
| Biological Sciences | 7.09 | 2.25 | 15.43 | 6.51 | 1.82 |
| Health Sciences | 12.61 | 2.80 | 14.38 | 11.37 | 2.02 |
| Exact and Earth Sciences | 6.56 | 1.57 | 11.33 | 7.53 | 2.27 |
| Human Sciences | 5.97 | 1.85 | 8.10 | 9.40 | 1.92 |
| Applied Social Sciences | 3.79 | 1.26 | 10.14 | 1.33 | 1.05 |
| Engineering | 3.51 | 1.22 | 11.80 | 3.57 | 2.05 |
| Linguistics, Languages and Arts | 3.29 | 1.29 | 3.70 | NaN | 1.00 |
| Other | 10.19 | 2.05 | 14.53 | 8.98 | 1.62 |

**Table 1.** Averages of mentions of the areas of study by type of source
Source: research data (2017)

The first point to be noticed is the good performance of Mendeley, followed by Twitter and the low coverage of Facebook that despite being the most used social media in Brazil, it reached an average of mentions well below the others, except from blogs. The result is very similar to the studies that have analysed the difference of online attention among the data source. Both in the research by Priem, Piwowar and Hemminger (2012) that have analysed the altmetric data from the Public Library of Science (PLoS) collection, and by Alperin (2015) that verified the online attention from the Scientific Electronic Library Online (SciELO) collection from Latin America countries, a higher coverage of data from Mendeley was identified, with a reading average well above the mentions on Twitter, which had well above averages compared to the posts on Facebook.

It is interesting to notice the difference of performance of the areas of study according to the type of source. Mendeley, for instance, registers higher values in the Biological Sciences (15.43) – a similar result to Alperin (2015) study – while the Health Sciences have higher coverage on Twitter (12.61) and on Facebook (2.80). Another highlight is the performance of press media with averages that provide them a higher coverage of Health Sciences (11.37), followed by Human Sciences (9.40) and no coverage in the area of Linguistics, Languages and Arts. The areas with a better performance in blogs posts are the Exact and Earth Sciences (2.27) and Engineering (2.05).

According to Alperin (2015), these fluctuations among areas in terms of coverage levels indicate the different uses of each type of source, potentially indicating that different audiences are being reached by each discipline or different practices in relation to the social media and reference managers.

**4 Conclusion**

The Brazilian production in a specific area of study authored by the productivity scholarship holders from CNPq reflects the research agenda and contitutes a body of privileged knowledge about the representation of this area in the country, expressing commitments with theoretical, methodological and thinking currents about its historicity, perfomance and trend.

This study aimed to evaluate this production from the perspective of its social impact, measured, not by citations, but by the online attention that such research arouse inside and outside the scientific community when mentioned in social media. Qualitative studies are needed to ascertain contents and contexts of these mentions and to improve the altmetric picture of the analyzed production.

The low result found in the indexes and averages of mentions, in a general way, according to the area of study or the type of source, make us question whether it is a matter of a non-culture of sharing scientific production in social media or if it indicates, as pointed out by Alperin (2015), a matter of altmetric data limitation and coverage for the countries of Latin America.